# Advanced Mid-Infrared Plasmonic Waveguides for On-Chip Integrated Photonics

*Mauro David\*, Davide Disnan, Elena Arigliani, Anna Lardschneider, Georg Marschick, Hanh T. Hoang, Hermann Detz, Bernhard Lendl, Ulrich Schmid, Gottfried Strasser, Borislav Hinkov\**

M. David, E. Arigliani, A. Lardschneider, G. Marschick, H. T. Hoang, H. Detz, G. Strasser, B. Hinkov
Institute of Solid State Electronics, TU Wien
Gußhausstrasse 25-25a, 1040 Vienna, Austria
Email Address: *mauro.david@tuwien.ac.at*
Email Address: *borislav.hinkov@tuwien.ac.at*

D. Disnan, U. Schmid
Institute of Sensor and Actuator Systems, TU Wien,
Gußhausstrasse 27-29, 1040 Vienna, Austria

H. Detz
CEITEC, Brno University of Technology Žerotínovo nám. 617/9, 601 77 Brno, Czech Republic

B. Lendl
Institute of Chemical Technologies and Analytics
Getreidemarkt 9, 1060 Vienna, Austria

***Abstract-*** *Long-wave infrared (LWIR, 8-14 µm) photonics is a rapidly growing research field within the mid-IR with applications in molecular spectroscopy and optical free-space communication. LWIR-applications are often addressed using rather bulky tabletop-sized free-space optical systems, preventing advanced photonic applications such as rapid-time-scale experiments. Here, device miniaturization into photonic integrated circuits (PICs) with maintained optical capabilities is key to revolutionize mid-IR photonics. Sub-wavelength mode confinement in plasmonic structures enabled such miniaturization approaches in the visible-to-near-IR spectral range. However, adopting plasmonics for the LWIR needs suitable low-loss and -dispersion materials with compatible integration strategies to existing mid-IR technology. In this work we further unlock the field of LWIR/mid-IR PICs, by combining photolithographic patterning of organic polymers with dielectric-loaded surface plasmon polariton (DLSPP) waveguides. In particular, polyethylene shows favorable optical properties, including low refractive index and broad transparency between ~2-200 µm. We investigate the whole value chain, including design, fabrication, and characterization of polyethylene-based DLSPP waveguides and demonstrate their first-time plasmonic operation and mode guiding capabilities along s-bend structures. Low bending losses of ~1.3 dB and straight-section propagation lengths of ~1 mm, pave the way for unprecedented, complex on-chip mid-IR photonic devices. Moreover, DLSPPs allow full control of the mode parameters (propagation length and guiding capabilities) for precisely addressing advanced sensing and telecommunication applications with chip-scale devices.*

***Keywords*:** plasmonic waveguides, mid-infrared integrated photonics, polyethylene, polymers, long-wave infrared

## 1. Introduction

The long-wave infrared (LWIR) spectral region (8-14 µm) is a portion of the electromagnetic spectrum covering many applications of significant interest, ranging from high-performance spectroscopy[1–4] to optical free-space communication[5]. In particular, chip-scale LWIR photonics has enormous potential for realizing unprecedented miniaturized devices. By specifically tailoring their characteristics, novel device concepts could be unlocked addressing, e.g., real-time reaction monitoring of (bio-)chemical samples[6–8], *in situ* health monitoring using (medical) wearables[3], and ultra-fast optical data communication[9–11]. While today's most promising LWIR coherent light source, the Quantum Cascade Laser (QCL), is commercially available, offering high power, room-temperature, and continuous-wave operation[12–14], its monolithic counterpart, the QC Detector (QCD), is currently still under development in this wavelength range[15,16]. In a first integration step, both can be combined into specifically tailored devices with same-wavelength emission and detection from the same active region[13,17,18], paving the way for a new class of monolithic photonic integrated circuits (PICs)[6,19]. However, in contrast to near-IR PICs with their high degree of maturity[20], mid-IR (2-20 µm) PICs are still in the early stages of development[21], and implementing existing integrated photonics solutions remains a challenging task[21,22]. It includes the requirement for new materials[23] and advanced fabrication schemes for developing low-loss and low-dispersion LWIR passive photonic structures on the chip-level. Especially, concepts that allow efficient mode guiding and directing are currently lacking as crucial elements for fully integrated devices with versatile capabilities and complex functionalities. One main reason for this is that the vast majority of commonly used materials become lossy in the LWIR[22], and scaling geometrical parameters with the wavelength becomes unfavorable, preventing their use for monolithic integration[24]. This is true, for example, for dielectric waveguides based on Ge[25,26], Si[27,28], halides[29], chalcogenides[30] or heavy metal oxides[31]. Nevertheless, recent work made noticeable progress by integrating InGaAs/InP passive dielectric waveguides with InP-based mid-IR QCLs[24,32]. However, this integration method requires a complex post-regrowth fabrication scheme that can potentially decrease fabrication yield[33]. Another promising, and even simpler route for developing monolithic mid-IR PICs, is presented by the use of dielectric-loaded surface plasmon polariton (DLSPP) waveguides[34,35]. They combine several advantages including tailorable propagation lengths and sub-wavelength mode confinement for further device miniaturization, together with compatible, available integration approaches.

Plasmonic waveguides have several benefits over traditional dielectric waveguides, such as stronger mode confinement and higher field intensity[36]. They also exhibit enhanced light-matter interactions, making them promising candidates for applications in sensing[37], spectroscopy[38], and nonlinear optics[39]. Regarding fabrication, surface plasmon polariton (SPP) waveguide-based devices benefit from using the same metal(s) for carrying optical signals as typically used for the electrical contacts of the device (e.g., Au). This simplifies their monolithic integration significantly[40]. Indeed, DLSPP waveguides based on 200-300 nm thick $Si_3N_4$ stripes on Au were already successfully demonstrated at ~6.2 µm as linear optical couplers between QCL and QCD in novel mid-IR sensors[6,19,41]. However, the application of $Si_3N_4$ as a confinement layer is limited in the LWIR due to strong absorptions above 7 µm[42]. Moreover, the effective mode index of the presented structures ($N_{eff} \approx 1.008$) is too low to provide efficient on-chip guiding, which is a necessity for more complex chip-scale optical circuits with advanced capabilities. Hereby, being able to guide a LWIR optical signal on a pre-defined path, potentially allows a much higher degree of control on the direction of the mode, as compared to what could be achieved by controlling the light emission direction of individual discrete elements like QCLs[43].

Despite providing broadband and low-loss transmission throughout the entire mid-IR spectrum (2-14 µm), similar limitations emerge for semiconductor-loaded SPP (SLSPP) waveguides based on, e.g., previously developed Ge/Au architectures[44]. Due to the broad transparency of Germanium, it is possible to increase the $N_{eff}$ by appropriately tailoring the ridge cross-section. However, this comes at the expense of a significantly lower propagation length[44]. Therefore, new materials are needed for the LWIR spectral range, providing high transparency and low refractive index (RI), thus enabling similar performances as for their near-IR counterparts.

Organic polymers have always been attractive candidates for such applications in PICs due to their low cost, biocompatibility, mechanical flexibility, low optical losses, ease of tunability, and compatibility with established nanofabrication processes[45,46]. Unfortunately, similar to many inorganic materials, the most commonly used polymers with excellent optical properties in the near-IR, such as PMMA and SU-8, cannot be used at longer wavelengths since they strongly absorb in the LWIR region[47,48]. In contrast, polyethylene (PE) has recently been proposed as a promising material for mid-IR integrated photonics applications[49] based on its broad optical transparency (~2-200 µm)[49,50]. Methods to fabricate PE-based waveguides or antennas have already been reported for THz (>20 µm) applications[52,51,53]. However, the lithographic surface patterning of PE films with micrometer resolution, necessary for the fabrication of high-precision chip-scale optical interconnects, was not reported so far. Moreover, the fundamental demonstration of LWIR plasmons in PE-based DLSPP waveguides was also not yet shown in literature.

In this work, we employ highly transparent PE films synthesized via spin-coating as a ridge material for LWIR DLSPP waveguides. We use finite element simulations performed with the commercial software tool COMSOL to design and optimize various ridge geometries. We then fabricate the plasmonic devices by a novel UV-photolithographic process through the definition of an inorganic hard mask and $O_2$ plasma etching. The device's optical performance is experimentally studied, and the propagation losses are extracted using the cut-back method. Our approach demonstrates the possibility of patterning PE to a wide variety of ridge geometries via a straightforward photolithography fabrication process. It enables full control of waveguide structures allowing to accurately tailor them to various dedicated applications. Besides demonstrating a new SPP waveguide system for the LWIR, we also unlock a new material platform for mid-to-far-IR integrated photonics. With this, we open an avenue for a new class of complex monolithic photonic devices and networks, exploiting the combination of PE-based waveguide structures with QC technology for the mid-IR and beyond.

## 2. Results and discussion

### 2.1. Simulation and Design of the Plasmonic Structures

The first step in the device design is based on performing simulations of various device structures conducted using the eigenmode solver of COMSOL Multiphysics. As the accuracy of the calculations strongly depends on the optical parameters of the materials involved, we optimized the thin film fabrication technique to achieve high-quality layers with optical parameters similar to the bulk materials. Further details on the fabrication and the RI data of the PE thin films are available elsewhere [49]. The simulations were carried out for a target wavelength of 9.26 µm, which is in one of the spectral windows suitable for free-space optical telecommunication[5]. The plasmonic waveguide architecture under investigation is the DLSPP waveguide, shown in Figure 1a. It consists of a dielectric ridge on top of a metal (Au) layer. Figure 1b shows the electric field confinement ($|E_y|$) for a PE ridge of 3.6 x 3.6 µm². For these design parameters, the mode is localized in the dielectric core of the waveguide at the metal-dielectric interface, showing good propagation length ($L_P$ ~ 455 µm) with an effective mode index of $N_{eff}$=1.20. This demonstrates the promising performance characteristics of this material system for combining long-range $L_P$ with high enough $N_{eff}$ for potential optical mode guiding along the chip surface.

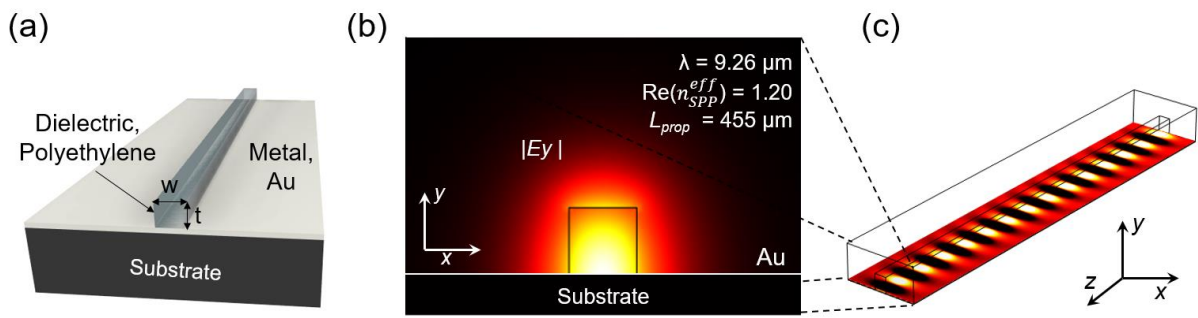

Figure 1. a) Cross section of the DLSPP waveguide. b) $|E_y|$ 2D field profile of the $TM_{00}$ SPP mode for a 3.6 x 3.6 µm² DLSPP waveguide at λ = 9.26 µm. c) Top view of the 3D simulation of the waveguide.

Figure 2 shows the impact of the ridge dimensions on different figures of merits. Figure 2a displays the simulated effective mode index for different thicknesses, plotted over the ridge width. The figure highlights the value of $N_{eff} = 1.1$, yielding the minimum refractive index contrast to achieve effective mode-guiding along waveguide bends in a surrounding cladding of air. For increasing ridge thicknesses, the $N_{eff}$ monotonously increases and enables better mode guiding capabilities. Figure 2b presents equivalent simulations, showing the impact of varying ridge dimensions on the propagation length $L_P$. Overall, Figure 2 demonstrates the typical trade-off for plasmonic waveguides, where an increased $N_{eff}$ leads to increased mode confinement in the dielectric ridge at the expense of higher waveguide losses, i.e., lower $L_P$. For thin ridge geometries (t < 1.2 µm), the propagation length is in the order of 1 mm, however, $N_{eff}$ is too low to provide efficient mode guiding. Increasing the thickness of the ridge enlarges $N_{eff}$ and provides the necessary condition for light guiding, however at the expense of the propagation length.

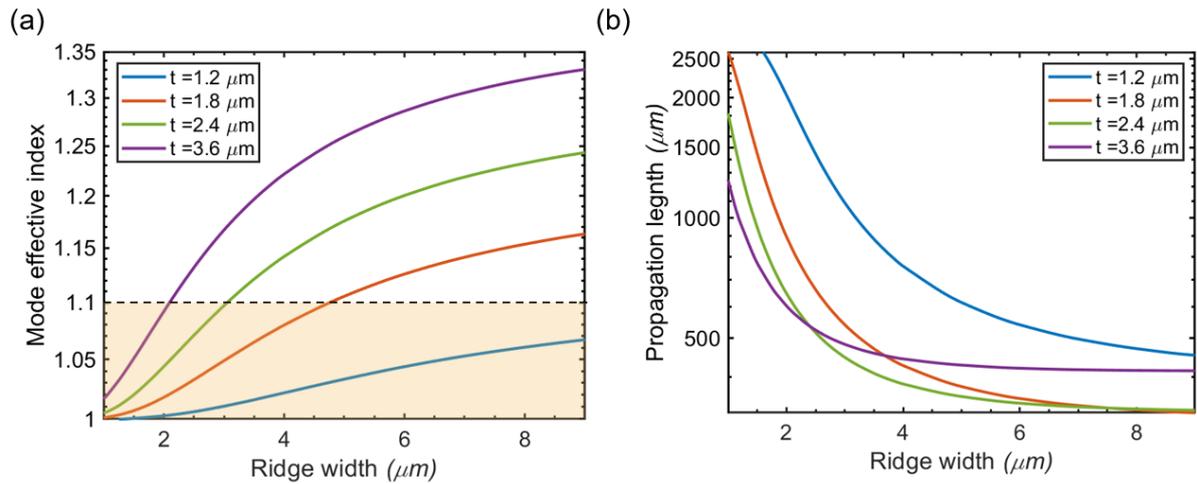

Figure 2. a) Mode effective index $N_{eff}$ for different ridge dimensions. b) Propagation length $L_p$ for different ridge dimensions.

These results show the similarity in terms of performance when comparing PE-based DLSPP waveguides with previously reported numerical studies for PMMA-based DLSPP waveguides at telecom frequencies[34].

The above analysis of the geometrical factors performed via numerical simulations highlights specific ridge dimensions that are highly favorable for integrated photonics applications. In particular, Figure 3a shows that there is an optimal ridge width (in the range of 2-5 µm) to minimize mode size, here expressed in the form of normalized mode area $A_{eff}/A_0$. As in the case of PMMA-based DLSPP waveguides at telecom frequencies[54], for smaller-than-optimal ridge widths, the mode is pushed outside of the dielectric ridge, increasing its extension. For larger ridge widths, the mode will occupy all the available dielectric space leading again to a lower degree of mode confinement (see also inset in Figure 7b). Remarkably, for PE ridges with optimal geometries, the transmission performance can be highly comparable to the well-known PMMA-based DLSPP waveguides at telecom frequencies. This is shown in Figure 3b, where we simulated the transmission $T_b = I_b/I_a$ (see inset in Figure 3b) for a 90° bend over a waveguide section of circular shape. For both, PE at $\lambda = 9.26$ µm and PMMA at $\lambda = 1.55$ µm, the reported transmission is displayed as a function of normalized radius $R/\lambda$[55]. The plot shows that the optical properties of both polymers at their respective operational wavelength are very similar and that by simply scaling the geometry, it is possible to reproduce the plasmonic performance of PMMA at telecom frequencies in PE for the LWIR spectral region. Using these simulation results as design guidelines, we proceeded with the fabrication of the plasmonic devices.

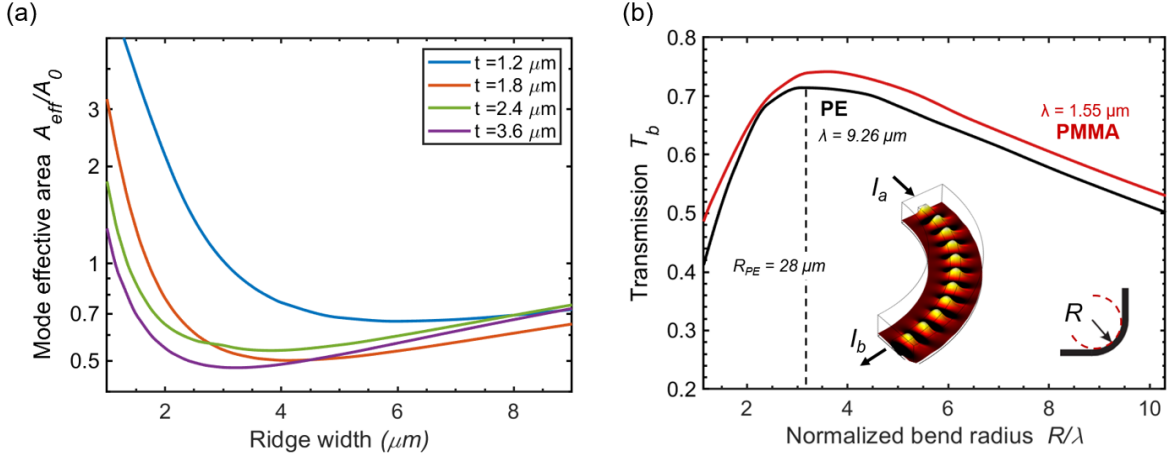

Figure 3. a) Influence of the polyethylene ridge geometries on the effective mode area ($A_{eff}/A_0$). b) Comparison of the transmission ($I_b/I_a$) of a 90° polyethylene DLSPP waveguide bend (cross-section of 3.6 x 3.6 µm²) at $\lambda$ = 9.26 µm and a PMMA DLSPP waveguide (cross-section of 600 x 600 nm²) at $\lambda$ = 1.55 µm, as a function of its normalized radius ($R/\lambda$). Inset: 3D modeling of the polyethylene waveguide mode.

## 2.2. Fabrication of the Plasmonic Devices

Gold is one of the most commonly used metals for electrical contacts of mid-to-far infrared photonic devices, making it a suitable choice for plasmonic layers, as discussed previously. Another important benefit of employing metal-based plasmonics is the independency from the substrate material, allowing a much more flexible integration. In other words, it provides greater flexibility in the choice of substrate material for the photonic device. Indeed, the penetration depth of the evanescent field into the metal for LWIR frequencies amounts to only around 13 nm. This enabled the use of Si substrates coated with a 100 nm plasmonic Au layer, while QC technology-based material systems such as InP or GaAs can also be used. The PE films used as core material for the DLSPP waveguides were deposited by spin-coating from a solution of concentration wt=10%, resulting in films of 5.0 µm thickness. For the targeted wavelength of 9.26 µm, the real and imaginary parts of the refractive index of the PE films were measured with ellipsometry[49] and found to be 1.48 and 0.0001, respectively. Figure 4 shows the schematic flowchart of the fabrication steps of the plasmonic devices. As PE is not inherently an UV-sensitive polymer[56], we used a similar approach that has been used for organic PVDF-based MEMS[57], i.e., by definition of an inorganic hard mask. In the case of an Au hard mask (Figure 4a), the lithographic process must be repeated a second time in order to safely remove the top Au-layer without damaging the one at the bottom. In following attempts, the fabrication technique was further optimized using a Cr hard mask (Figure 4b), which significantly reduced the number of "mouse-bite-type" defects at the bottom of the waveguide because the Cr etchant does not attack the bottom Au layer. As additional advantage, this also avoids the need for a second lithography step. Further details on the film and ridge fabrication can be found in the Experimental Section. After the device fabrication, the chips were finalized by cleaving the plasmonic waveguides into different lengths (ranging from 1 to 2.3 mm) for cut-back measurements.

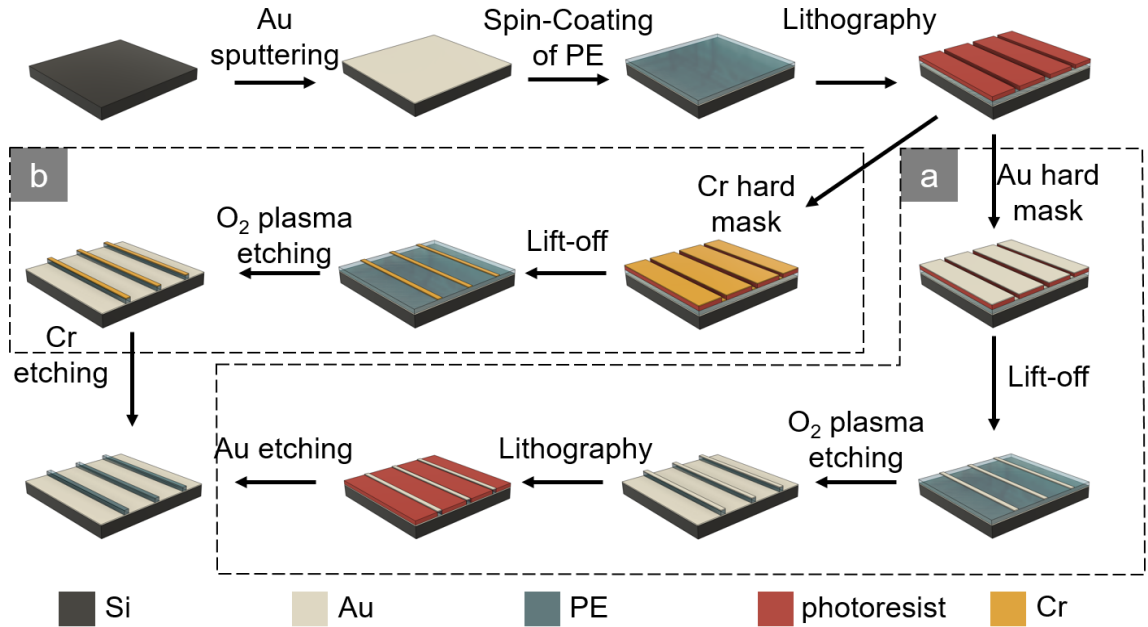

Figure 4. Schematic flow chart of the chip fabrication process, a) using an Au hard mask and b) using a Cr hard mask.

The surface topography of the spin-coated thin films was then characterized using atomic force microscope (AFM). Figure 5 shows the root-mean-square (rms) area roughness of the PE layer after spin-coating. The rms roughness of the PE layer was found to be 10.4 nm, suggesting a relatively smooth surface as result of the spin-coating process. Notably, no significant difference in surface topography was observed between the top of the fabricated devices and the spin-coated films prior to surface patterning ($S_q$ = 10.7 nm).

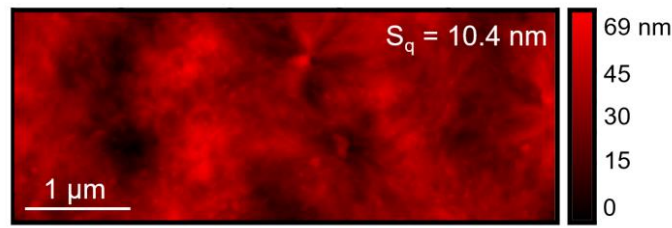

Figure 5. Typical AFM image of the fabricated polyethylene films right after spin-coating. Rms area roughness amounts to $S_q$ = 10.4 nm.

Following fabrication, the structures were analyzed with scanning electron microscope (SEM) to evaluate their physical characteristics (Figure 6). The SEM images revealed that the sidewalls of the waveguide exhibited a certain degree of roughness that can potentially lead to increased scattering losses and reduced efficiency of the device. However, the upper waveguide edge appeared to be relatively flat, indicating that the fabrication process was able to achieve a good level of precision.

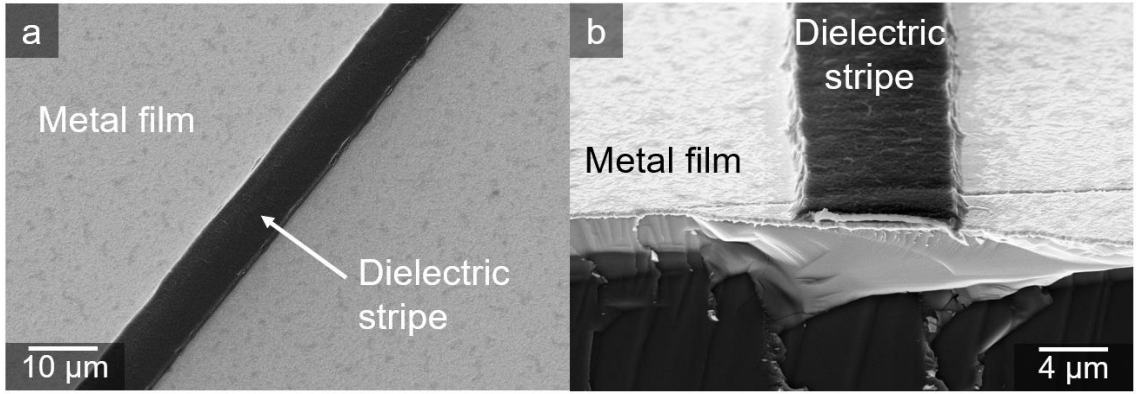

Figure 6. Scanning electron microscope (SEM) image of a DLSPP PE-waveguide structure. a) Plasmonic waveguide and b) Close-up of the 8 µm wide and 1 µm high PE stripe.

Several geometries were investigated varying the widths from 3 to 20 µm. The devices were etched using a plasma etching system (STS 320PC–without implemented temperature control). After the plasma etching process, we observed an inhomogeneous decrease in thickness after each etching step. As the polymer has a relatively low melting point (110°C-135°C), we attribute this to the uncontrolled temperature reached in the chamber combined with the low pressure during the etching process. In particular, devices fabricated with a 20 nm Cr-hard mask, as opposed to 100 nm of Au, exhibited a minor decrease in thickness, indicating that the polymeric films potentially reached the softening point (~110°C) during the etching process and were affected by the weight of the hard mask. However, because of these unexpected changes in thicknesses, we obtained devices with wider variety of geometrical parameters, allowing us to perform a much more general and complete comparison of the experimental results with the theoretically predicted ones.

### 2.3. Transmission Measurements

The optical performance of the plasmonic devices was determined by transmission measurements using a custom-built measurement setup[44]. The propagation losses were determined by the effective cut-back method for each fabricated waveguide geometry[58–61]. The summary of the results in terms of waveguide and coupling losses is shown in Figure 7a. The fabricated devices were modeled in COMSOL, and the losses and normalized area results are displayed as dashed lines. The deviation between the measured data of Figure 7a with the model can be attributed to fabrication imperfections leading to differences in device performance. Overall the measurements show excellent agreement with simulations. Figure 7b shows the corresponding mode cross-sections for the different dimensions.

In SPP waveguides, the supported vertical and lateral mode confinement is strongly dependent on geometrical factors, as depicted in Figure 7b. This relationship is experimentally validated by comparing measured coupling losses with the simulated normalized effective mode area ($A_{eff}/A_0$). In an end-fire coupling experiment, coupling losses occur when the input light mode is not perfectly matched to the mode size supported by the receiving optical component. Hence, larger effective mode areas lead to higher coupling losses (for lenses with small focal spot sizes, as in this case), since the spatial overlap between the lens and waveguide mode is reduced. Conversely, smaller effective mode areas result in lower coupling losses due to the higher spatial overlap between the two components. In conclusion, our measurements fit well with the simulation results.

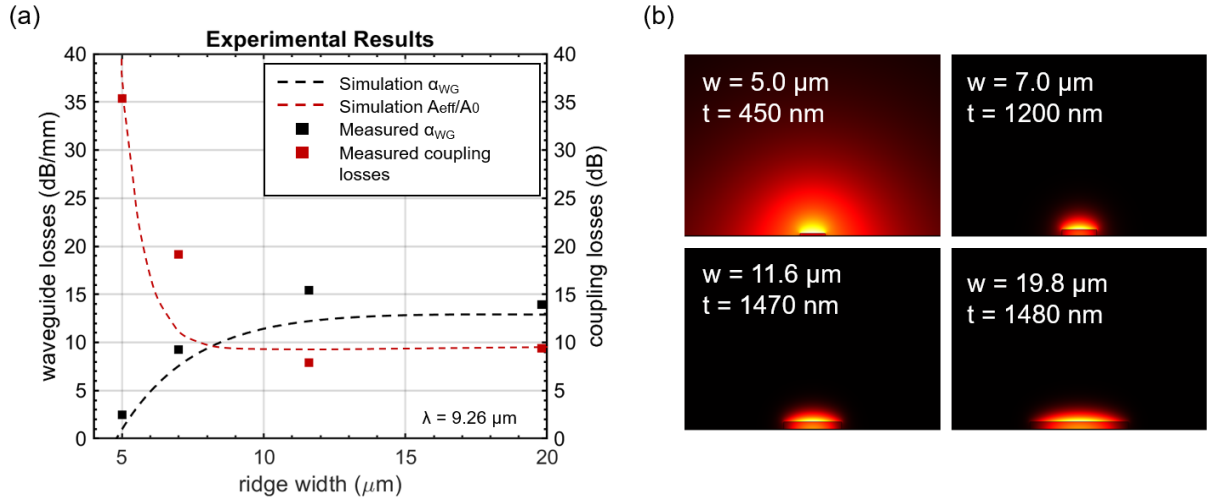

**Figure 7.** a) Summary of measured data in terms of waveguide losses (left) and coupling losses (right) extracted from the effective cut-back method. The square points represent the losses extracted with the cut-back method per each waveguide width. Dashed lines represent the exponential fit of the results obtained from the simulations. b) Mode-profile simulations of the fabricated DLSPP waveguides, indicating their respective width (w) and thickness (t) measured with a profilometer.

The lower RI of this polymer allows for the first-time replicating a similar DLSPP waveguide performance in the LWIR as compared to PMMA at telecom wavelengths. In contrast, the implementation of transparent high-RI materials, such as Si or Ge, requires the use of much smaller geometries (width and thickness approximately 1 x 1 µm$^2$) to achieve similar routing performances. However, in the latter case, achieving such low-loss propagation is not feasible due to the higher degree of confinement that amplifies the losses of the optical mode because of the lossy Au layer. Moreover, their fabrication and integration with active devices on the same chip can be challenging with such small dimensions.

To experimentally investigate the bend losses of the plasmonic waveguides, we fabricated 5 µm wide and 3 µm thick waveguide structures (Figure 4b) in the form of double S-bends (Figure 8a, between "A" and "B"). The design of the bend was optimized based on the simulations reported in Figure 3b, targeting a bend radius of 28 µm and ensuring adiabatic evolution of the mode throughout the bend. Two devices were fabricated with bend offsets (i.e., the difference between the two arms) of $d_1 = 50$ and $d_2 = 75$ µm. The results are summarized in Table 1. The experimental and theoretical transmission values were found to be in good agreement, as shown in Figure 8b. They demonstrate the significant potential of PE as a promising material for DLSPP waveguides for mid-to-long infrared integrated photonic applications.

Table 1. Table reporting the simulated and measured S-bend losses (for a 106 µm long S-bend). The scattering losses due to fabrication defects are calculated by subtracting the simulated losses from the measured losses.

| Offset [µm] | Simulation [dB] | S-bend losses [dB] | Scattering (defects) [dB] | Band losses [dB] |
|---|---|---|---|---|
| 50 | 2.30 | 3.18 | 0.88 | 1.31 |
| 75 | 3.45 | 5.22 | 1.77 | 2.33 |

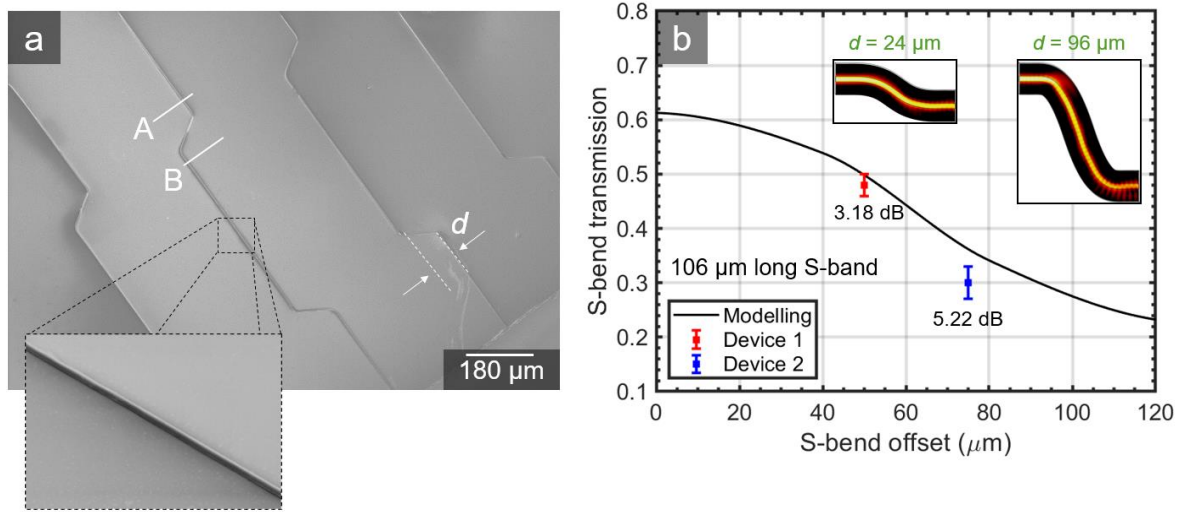

Figure 8. a) SEM images of the S-bend structures (width w = 5 µm and thickness t = 3 µm) with an offset of d = 75 µm. b) S-bend transmission dependence on the offset distance between the two arms (section A-B) obtained by numerical modeling and measurements. (inset) top-view of the 3D simulations for offsets of d = 24 µm and d = 96 µm.

The findings of this study carry significant implications for the development of organic-based biocompatible photonic integration concepts utilizing plasmonic or other photonic waveguide designs for the LWIR, including traditional dielectric or photonic crystals waveguides and optical fibres. Therefore, the demonstrated low-loss properties of the PE utilized in this study and its successful fabrication present an highly promising path toward realizing fully integrated and complex photonic systems. Furthermore, the success of the UV-lithography process used to pattern PE films offers new possibilities for the realization of other low-loss waveguide architectures in the mid- to far-IR range. In addition, the same fabrication routine can be used to create transparent microfluidic chips with micrometer feature size, offering a new platform for on-chip fluidic transport and optical sensing. This approach could be used to develop fully-integrated lab-on-a-chip systems that are largely transparent in both, the visible and the mid-IR, while unlocking a wide range of biological and chemical sensing applications. Additionally, it can enable the development of novel low-loss waveguide architectures for applications beyond the LWIR and in the THz range.

## 3. Conclusion

We have realized the first demonstration of PE-based plasmonic waveguides in the LWIR. This opens the much-needed pathway for realizing more complex LWIR monolithic PICs and networks on the chip-scale, with a wide variety of tailorable properties. After spin-coating of the organic PE thin films, a suitable fabrication process using plasma-etching and a metallic hard mask was developed to pattern the plasmonic structures. Transmission measurements show good agreement with simulations in terms of waveguide losses. From the experiments, the measured waveguide propagation losses of the fabricated devices lie between 0.00248 and 0.01545 dB $\mu m^{-1}$ and bend losses are ~1.3 – 2.3 dB. These values can be further improved by optimizing the fabrication process. The promising material properties of this polymer will open new pathways for organic-based integrated photonic devices. They are not limited to plasmonic concepts and offer the coverage of a broad spectral range, from mid-to-far infrared frequencies.

## 4. Experimental section

*Simulation and Design of the Plasmonic Devices:* To carefully investigate the waveguiding properties of the PE-based DLSPP waveguides, we performed numerical simulations with the well-established optical mode solver of COMSOL Multiphysics. Based on the obtained refractive index data from our previous ellipsometry measurements available in literature[49], we carried out numerical mode analysis for a target wavelength of 9.26 µm. The simulation results served as design guidelines for the fabrication of the devices and for later estimation of the scattering losses caused by fabrication imperfections, such as structural defects and surface roughness. The figures of merit used as design parameters are the effective mode index $N_{eff}$, propagation length $L_p$, and normalized area $A_{eff}/A_0$. The propagation length is defined as $L_p = 1/(2\alpha)$, where $\alpha$ is the attenuation constant. The normalized area is defined as the ratio between effective mode area $A_{eff}$ and the diffraction limit $A_0 = \lambda^2/4$, where,

$$A_{eff} = \frac{1}{Max\{W(r)\}} \int_{A_\infty} W(r) \, dA \qquad (1)$$

with the energy density $W(r)$ at position r,

$$W(r) = \frac{1}{2} Re\left\{\frac{d[\omega\varepsilon(r)]}{d\omega}\right\} |E(r)|^2 + \frac{1}{2} \mu_0 |H(r)|^2 \qquad (2)$$

*Synthesis of the Polyethylene films:* Amorphous PE films were spin-coated in a humidity- and temperature-controlled glove box. The films were synthesized from linear low-density PE powder (Sigma-Aldrich, $\rho$ = 0.906 g mL$^{-1}$, average $M_w$ ~ 35000 and $M_n$ ~ 7700) dispersed in toluene (Sigma-Aldrich, anhydrous, purity 99.8%) with a mass portion of 10%. The spin coating was conducted from a hot solution (100°C) at a rotation speed of 1000 rpm for 35 s, with 5 s of acceleration, using the dynamic dispense method. The samples were pre-heated to 120°C on a hotplate before spin-coating and post-baked at 120°C for 15 minutes right after spin-coating. Details about the synthesis of the films, as well as their optical and structural characterization, can be found in our previous work[49].

*Fabrication of Plasmonic Devices:* Double-side polished 2" Si (100) wafers with 275 µm thickness were used as substrates. After cleaving, the 11 x 11 mm² samples were ultrasonically cleaned in an acetone bath before the deposition of 8 nm of Ti and 100 nm of Au using a standard RF-sputtering machine (LS 320S, Von Ardenne). Subsequently, the polymer was spin-coated on the Au-coated samples targeting a thickness of 5 µm. For patterning, AZ5214E photoresist was spin-coated directly on top of the polymer at 4000 rpm for 30 s, followed by baking at 100°C for 1 min, resulting in a 1.5 µm thick resist layer. The resist was then patterned using an UV mask aligner (Karl Suess, MJB4) system for contact lithography with 4 s of exposure time, followed by 37 s of development in AZ726MIF. To transfer the pattern from the resist to the polymer, the sample was covered by a 100 nm Au layer (or 20 nm of Cr) deposited by e-beam evaporation. After a lift-off procedure (30 min in acetone), the samples were etched using a plasma etching system (STS 320PC –no temperature control). The etching recipe used is based on a standard oxygen plasma (power: 150 W, O2 flow rate: 49.5 sccm, pressure: 150 mTorr, time: 4 cycles of 5 min of etching with 10 min. for cooldown) that can be further optimized to control the etch rate, sidewall roughness and reduce redeposition on the substrate. The resulting etch rate was 250 nm min$^{-1}$. The surface morphology of the fabricated devices is shown in Figure 6. After the etching, a second lithography process is performed, similarly as described above to remove the Au hard mask. Finally, the Au hard mask is etched using KI/I$_2$ gold etchant for 30 s, and the resist is subsequently removed in an acetone bath. The fabrication process flow is shown in Figure 4, and can be improved by the use of a Cr hard mask, as described in the Results and Discussion section. The final devices were prepared for optical characterization by cleaving it into different lengths.

*Transmission Measurements:* The devices were characterized in an end-fire coupling geometry using a custom-made setup consisting of a 2-lens geometry (C037TME-F, Thorlabs). The device under test was placed on a piezoelectric transition stage. The waveguides were tested at a wavelength of $\lambda$ = 9.26 µm using a HHL501 DFB-QCL. The laser light is directed to a mid-IR camera (Tau 2, Teledyne FLIR LLC, USA) for inspecting the beam profile and focused by a parabolic mirror onto a thermoelectrically cooled MCT detector (PVI-4TE-10.6, Vigo Systems S.A., Poland) through the use of a ZnSe beam splitter (BSW711, Thorlabs). To extract the waveguide losses, the effective cut-back technique is used[44,58–61]. The measurement procedure consists of inserting the waveguide between the lenses and carefully aligning the sample to maximize signal intensity. The reference measurement is taken prior to and after every reading, and the intensity of each waveguide is measured three times and averaged. The scattering losses from fabrication defects can then be calculated by subtracting the simulated propagation losses from the measured losses.

**Acknowledgments**

We acknowledge received funding from the EU Horizon 2020 Framework Program (project cFlow, No. 828893), B.H. acknowledges funding by the Austrian Science Fund FWF (M2485-N34). CzechNanoLab project LM2018110 funded by MEYS CR is gratefully acknowledged for the financial support of the measurements at CEITEC Nano Research Infrastructure. Fruitful discussions with Werner Schrenk, Aaron M. Andrews, and Benedikt Schwarz are gratefully acknowledged. We want to thank Alicja Dabrowska, Sandro Dal Cin, Cem I. Doganlar and Andreas Linzer for expert technical assistance.